**Surface Stability of SrNbO$_{3+\delta}$ Grown by Hybrid Molecular Beam Epitaxy**

Suresh Thapa[1], Sydney Provence[1], Steve M. Heald[2], Marcelo A. Kuroda[1], and Ryan B. Comes[1]

[1]Department of Physics, Auburn University, Auburn, AL 36830, USA
[2]Advanced Photon Source, Argonne National Laboratory, Argonne, IL 60439, USA

**Abstract:**
4d transition metal oxides have emerged as promising materials for numerous applications including high mobility electronics. SrNbO$_3$ is one such candidate material, serving as a good donor material in interfacial oxide systems and exhibiting high electron mobility in ultrathin films. However, its synthesis is challenging due to the metastable nature of the d$^1$ Nb$^{4+}$ cation and the limitations in the delivery of refractory Nb. To date, films have been grown primarily by pulsed laser deposition (PLD), but development of a means to grow and stabilize the material via molecular beam epitaxy (MBE) would enable studies of interfacial phenomena and multilayer structures that may be challenging by PLD. To that end, SrNbO$_3$ thin films were grown using hybrid MBE for the first time using a tris(diethalamido)(tert-butylimido) niobium precursor for Nb and an elemental Sr source on GdScO$_3$ substrates. Varying thicknesses of insulating SrHfO$_3$ capping layers were deposited using a hafnium tert-butoxide precursor for Hf on top of SrNbO$_3$ films to preserve the metastable surface. Grown films were transferred *in vacuo* for X-ray photoelectron spectroscopy to quantify elemental composition, density of states at the Fermi energy, and Nb oxidation state. *Ex situ* studies by X-ray absorption near edge spectra illustrates the SrHfO$_3$ capping plays an important role in preserving the Nb 4d$^1$ metastable charge state in atmospheric conditions.

**Introduction:**

For two decades, the study of two dimensional electron gas (2DEGs) in complex metal oxides has rapidly increased after its observation at the LaAlO$_3$ (LAO)/SrTiO$_3$ (STO) interface [1]. This interesting phenomenon has driven ongoing research for two decades, opening the door for complex oxide interfaces as strong contenders for high carrier concentrations and high electron mobility. Oxide 2DEGs [2,3] offer unique opportunities compared to traditional semiconductor ones as they may exhibit strong spin-orbit coupling along with possibility of harnessing high carrier concentrations [citation?]. The search for a good donor oxide is a great challenge when building a high carrier concentration and mobility in oxides interface for high-speed electronics. Isoelectronic to the SrVO$_3$, SrNbO$_3$ (SNO) has a d$^1$ electronic configuration and a simple cubic perovskite structure with lattice parameter between 4.0 and 4.1 Å [4]. The band diagram of SNO reveals the Nb 4d t$_{2g}$ bands crossing the Fermi level, indicating its metallic nature, whereas the low work function makes SNO stand out as a suitable donor [5], as previously reported in density functional theory (DFT) studies on SNO/STO heterostructures [6]. SNO has also been discovered as the first plasmonic photocatalyst metallic oxide, which opens a door for new family of photocatalytic materials [7]. Studies have also examined the material for use as a plasmonic transparent conducting material due to the large band gap between the O 2p and Nb 4d bands [8–10]. Recent results have also shown that epitaxial strain in SNO films can break cubic symmetry to produce a semimetallic tetragonal phase with extremely high mobility and an observed Berry phase that make it promising for quantum materials applications [11]. Large linear magnetoresistances of ~10$^5$ % and mobilities of 80,000 cm$^2$/V-s have also recently been reported in SNO/STO heterostructures [12], suggesting that the material holds exceptional promise in topological and quantum materials research.

Despite the great potential of SNO for interfacial charge transfer and topological phenomena as well as its photocatalytic properties, limited work has been carried out in synthesis of pristine SNO thin films. There are numerous unanswered questions on basic properties of SNO, its surface stability, and its applications for interfacial engineering of emergent materials properties. Previous reports on SNO thin film growth have employed pulsed laser deposition (PLD) [10–15] and sputtering [8,9]. There are no reports of MBE synthesis of SNO, though $NbO_2$ [16] and Nb-doped STO [17] have been synthesized by MBE using an electron-beam evaporation source. In particular, previous work [14] examining the electronic band structure using *in situ* angle-resolved photoemission spectroscopy (ARPES) measurements provided insights into SNO thin films and verified the metallic character of SNO experimentally. However, this work also showed a significant excess of $Nb^{5+}$ atomic valence relative to the expected 4+ charge state, which suppressed the spectral weight of electronic states near the Fermi level. These results were attributed to surface $Nb^{5+}$ states due to either Sr vacancies or formation of $Nb_2O_5$. However, adsorption of excess oxygen is also an alternative mechanism to accommodate the conversion from metastable $Nb^{4+}$ to the stable $d^0$ $Nb^{5+}$ state through the formation of a $Sr_2Nb_2O_7$ phase [15,18,19]. Similar results have also been observed in $d^1$ rare earth titanates [20,21] and $SrVO_3$ [16] previously. Clearly, understanding and controlling the valence of Nb ions in SNO films is important to the development of interfacial heterostructures that leverage the $d^1$ electronic configuration and propensity of SNO to donate electrons to neighboring materials [11,12].

One of the most likely approaches to preserve metastable SNO is by capping films with an alternative oxide with greater atmospheric stability. To date, there have been no reports of suitable capping layers for SNO. However, such an approach has been employed for $DyTiO_3$ films via a $LaAlO_3$ cap, which significantly enhanced the ratio of metastable $Ti^{3+}$ vs $d^0$ $Ti^{4+}$ [21]. A perovskite oxide with a higher band gap, matching A site cation, and one that does not accept electrons from SNO would be a reasonable choice as a capping layer. $SrHfO_3$ (SHO) is well known for its large band gap of 6.07 eV [22] and is expected to have a band alignment that prevents electron transfer, [5] which makes it an ideal perovskite oxide for capping in this system. However, the delivery of both Nb and Hf is highly challenging due to their refractory natures. To overcome this discrepancy, the hybrid molecular beam epitaxy technique opens opportunities for Nb and Hf to be supplied as metal organic precursors.

In addition to PLD, molecular beam epitaxy (MBE) is well established for the synthesis of various metal oxides over the last three decades. However, repeatable growth of a complex metal oxide incorporating refractory metal cations with stoichiometric control by traditional MBE is extra challenging as they lack the adsorption-controlled growth window. In the past decade, hybrid MBE (hMBE) has been established as a state-of-the-art technique to grow complex metal oxides where refractory metals are supplied through a metal organic precursor and the A-site cation is supplied as a metallic source [23]. The hMBE technique is highly efficient in delivering low vapor pressure and refractory metals for repeatable growth of large varieties of high quality metal oxides such as $SrVO_3$ [24], $BaSnO_3$ [25], and $SrTiO_3$ (STO) [26,27] with improved stoichiometric control compared to the traditional MBE growth. This approach offers significant advantages over deposition with an electron-beam evaporator in an MBE, where achieving repeatable and stable growth conditions can be a challenge.

Selection of precursor becomes highly important in perovskite oxides growth involving refractory metals as they determine the delivery of B-site cation. Previous reports [28,29] on synthesis of $Nb_2O_5$ by atomic layer deposition (ALD) suggest that niobium ethoxide (NbOEt) and tris(diethalamido)(tert-butylimido) niobium (TDTBN) are the most suitable precursors as a Nb precursor. However, NbOEt decomposition rather than evaporation in its molecular form has been reported due to significantly lower vapor pressure of NbOEt compared to TDTBN [29]. TDTBN was liquid at room temperature and successfully evaporated for ALD at 65 °C using open boats due to the low vapor pressure of the precursor. Recent work has shown both thermal and plasma-enhanced ALD of $Nb_2O_5$ using TDTBN [30]. In the case of thermal ALD, significant amounts of $Nb^{4+}$ ions were observed in the films grown at high temperatures, likely because, unlike NbOEt, TDTBN contains no oxygen atoms in the molecule. The chemical structure of the molecule is shown in Figure 1. The lack of oxygen in the TDTBN precursor may afford greater control over the overall oxidation of the film because the delivery of molecular $O_2$ and oxygen plasma can be tuned to vary the oxygen chemical potential in the system over a wide range.

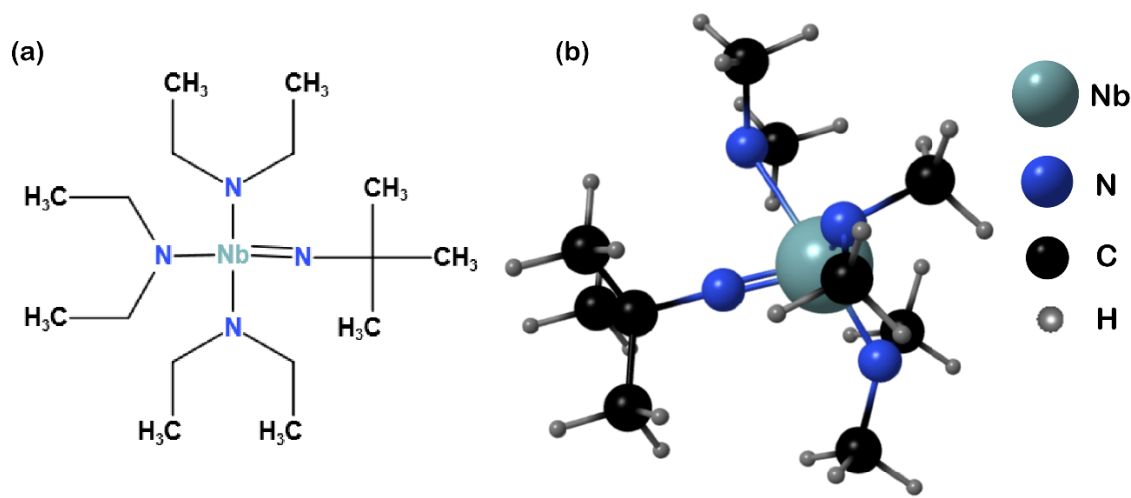

*Figure 1: (a) Structural formula and (b) ball and stick structure of tris(diethalamido)(tert-butylimido) niobium (TDTBN) precursor used for the hMBE growth of SNO epitaxial thin films.*

In this work, SNO band structure and partial density of states (pDOS) calculations are carried out by density functional theory (DFT). Metastable SNO films are grown on GdScO$_3$ substrate using hMBE. The surface of the film is monitored during the growth by reflection high energy electron diffraction (RHEED) and *in vacuo* X-ray photoelectron spectroscopy (XPS) is used to check the stoichiometry of the film. Possible over-oxidation of the surface or of the film is studied by analyzing the Nb 3d core level on uncapped and SHO capped samples. Correlating the valence band features further investigates the effectiveness of SHO capping in preserving Nb$^{4+}$ state. The surface roughness is analyzed using atomic force microscopy (AFM) topography. Further Nb K- edge was analyzed using X-ray near-edge spectroscopy (XANES) to check the effect of SHO capping thickness on Nb oxidation state.

**Methods:**
The electronic band structure of SNO was analyzed via DFT, using the Perdew-Burke-Ernzerhof (PBE) parameterization of the exchange-correlation energy [31]. No Hubbard U correction parameter was employed, consistent with previous models of the material [11,14]. Calculations employed projector-augmented-wave pseudopotentials [32] for the description of the atomic cores with cutoff energies of 40 Ry and 400 Ry for the wave functions and density, respectively. Cubic unit cells were fully relaxed until forces were smaller than $1.36 \times 10^{-7}$ eV/Å and band structure calculations were performed. The self-consistent potential was determined using an 8×8×8 Monkhorst-Pack mesh to sample the Brillouin zone. All calculations are carried out using the Quantum Espresso software [33].

Uncapped and SHO capped SNO thin films of identical thickness were grown on highly lattice matched (110) GdScO$_3$ (GSO) single crystal substrates (MTI Crystal) in a Mantis MBE reactor with a baseline pressure of ~10$^{-9}$ Torr. All the substrates were ultrasonically cleaned in acetone and isopropanol and dried with dry nitrogen gas. All the substrates were cleaned in oxygen in the MBE growth chamber by ramping to the 1000 °C growth temperature over ~1 hour as measured by thermocouple [26,34]. We estimate that the setpoint was ~150-200 °C higher than substrate surface temperature due to the absence of backside substrate metallization, resulting in GSO surface temperature between 800 and 850 °C. Cleaning and SNO (SHO) growths were performed in a background pressure of ~3×10$^{-6}$ Torr and ~3.6×10$^{-6}$ (~1×10$^{-5}$) Torr O$_2$ respectively. Calibration samples showed that the absorption of B-site cations takes a longer time compared to low flux of A-site cation. To control the exposure duration, alternate shutter deposition is applied to supply A and B-site cations in both SNO and SHO deposition. For the metalorganic sources, a pneumatic ALD valve (Swagelok 316L) is used to isolate the sources when they are not depositing and serves as the shutter for the deposition.

A series of uncapped and varying SHO thickness capped SNO samples were grown where Sr and tris(diethalamido)(tert-butylimido) niobium (TDTBN) fluxes were held fixed. The flux of SrO is measured by quartz crystal microbalance (QCM) in Å/sec. Strontium (99.99%, Sigma-Aldrich, USA) was supplied through low temperature effusion cell. The Sr flux was calibrated using QCM under oxygen environment with measured chamber pressure of $3\times10^{-6}$ Torr. Niobium was supplied through a gas source using the TDTBN metal-organic precursor (99.99%, Sigma-Aldrich, USA) from a bubbler connected to the growth chamber using the ALD pneumatic valve and a heated gas injector source (E-Science, USA). No carrier gas was used. The gas injector was held at 90 °C using a PID controller and the bubbler at 73 °C through external heating tape and a PID controller. While the vapor pressure was not measured directly, the chamber pressure showed only marginal changes (~1%) upon valve actuation, suggesting that the vapor pressure is in the range of 1-100 mTorr that is commonly employed for hMBE using a pressure-control feedback system [23]. This configuration can thus be thought of as a modified metal-organic evaporation source that replicates the capabilities of a low temperature effusion cell for organic molecules.

The growth chamber shroud walls were maintained at -30 °C via a closed loop chiller and low temperature fluid (Syltherm XLT, Dow Chemical) to reduce the background water vapor pressure from the dissociated TDTBN molecules. The chamber pressure measured by cold cathode gauge generally increases and reaches one unit higher at $10^{-6}$ Torr for the first 5-10 secs as the TDTBN supply valve is opened before stabilizing due to the dissociated and unreacted TDTBN injected into the system. $O_2$ flow is turned off for uncapped sample during cool down to minimize over-oxidation of the film or of the surface of the film. Right after the SNO deposition, the hafnium tert-butoxide (HTB) flux is calibrated and varying thickness of SHO capping layers were deposited turning on $O_2$ flow with measured chamber pressure of ~$1\times10^{-5}$ Torr $O_2$. Hafnium was supplied through a gas source using a metalorganic precursor, HTB (99.999%, Sigma-Aldrich, USA) from a bubbler connected to the growth chamber without carrier gas and controlled using a traditional pressure-control feedback loop [23,35]. Here we have presented four representative samples among all the samples including one uncapped and three capped annotated as thin (2 unit cells, ~0.8 nm), medium (3 unit cells, ~1.2 nm), and thick (4 unit cells, ~1.6 nm) capped.

*In situ* RHEED (Staib Instrument) was used to monitor the growth process and the quality of the film. Post-growth the samples were transferred from the MBE reactor to the PHI 5400 X-ray photoelectron spectroscopy (XPS) (Al $K\alpha$ X-ray source) system through ultra-high vacuum (UHV) transfer line. An electron neutralizer gun was applied to compensate charging of the insulating samples. The surface stoichiometry of all grown samples was characterized by analyzing core level and correlated with the valence band XPS spectra measured with base pressure of ~$1\times10^{-9}$ Torr. Analysis of the XPS data was performed using CasaXPS. To properly align all core level peaks, the valence band is fitted with a Fermi-Dirac function and all peaks are shifted with respect to the Fermi level because the charge neutralizer prevents direct determination of the Fermi level. A representative fit is shown in Figure S4 in the Supplemental Information. We assume in our analysis that electron forward focusing effects, commonly referred to as X-ray photoelectron diffraction, are negligible between samples. Though the probing depth for 95% of the signal is less than 5 nm which indicates these measurements are highly surface sensitive. A Rigaku Smartlab XRD equipped with a 4-circle goniometer and using Cu $K\alpha_1$ radiation line isolated with double bounce Ge (220) monochromator was used for $2\theta$-$\omega$ scans on the (002) reflection of the SNO samples. X-ray absorption near-edge spectroscopy (XANES) was performed at the Advanced Photon Source at Sector 20-BM in fluorescence mode for both in-plane (parallel) and out-of-plane (perpendicular) polarized x-rays at the Nb K edges.

## Results:

Figure 2(a) shows the cubic crystal structure of perovskite SNO where a niobium cation is octahedrally coordinated by oxygen ions. A lattice parameter of 4.02 Å was obtained using DFT calculations, which is agreement/disagreement with previous work [4]. The band structure and corresponding density of states (DOS) that provide context to interpret experimental results of SNO are shown in Figure 2(b) and 2(c), respectively. The metallic nature of SNO is observed with the Fermi level crossing conduction bands primarily formed by Nb 4d states, consistent with its $d^1$ electronic configuration. The O 2p bands make significant contributions to the band structure of SNO in the vicinity of the Fermi level. Fully occupied valence bands are composed mainly by the O 2p. The gap between the O 2p and Nb 4d bands is ~2.4 eV. These results agree with previous DFT reports [14].

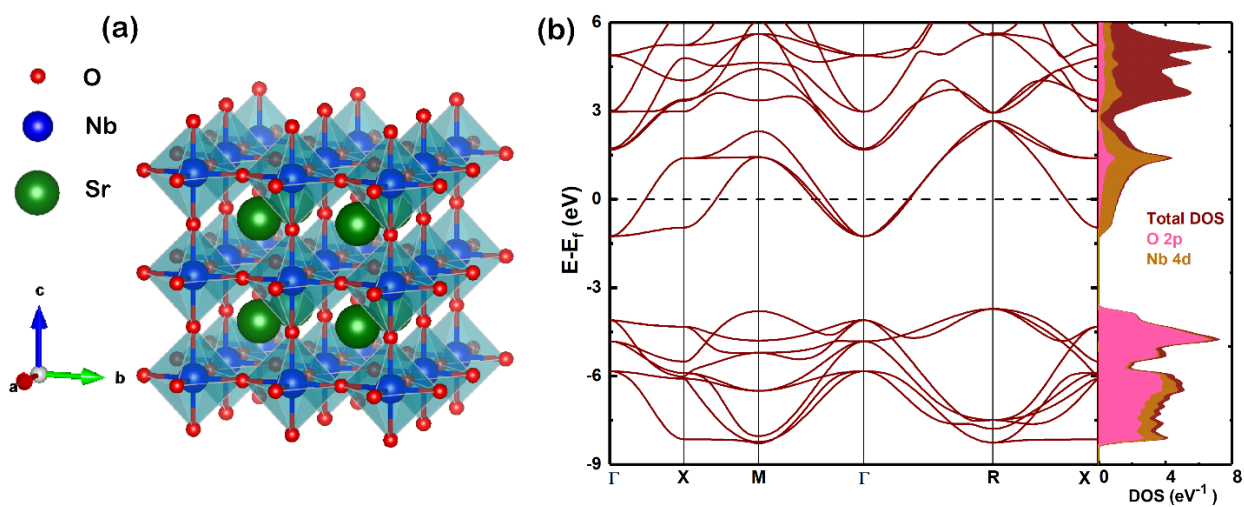

*Figure 2: (a) Cubic perovskite structure of SrNbO₃, (b) band dispersion and density of states (DOS) of SrNbO₃. Orbital decomposition onto localized atomic orbitals from Nb 4d and O 2p are shown in gold and pink, respectively.*

To test the efficacy of TDTBN as a precursor, SNO thin films were grown on GSO substrates by hMBE. Figure 3(a-b) shows the RHEED images for uncapped SNO film grown on GSO along the [110] azimuth immediately after growth at high temperature and after the sample is cooled down to room temperature. Intense diffraction spots as well as sharp Kikuchi bands in Figure 3(a) indicate that the sample is smooth with a well-crystalized surface film. However, the diffraction spots along with Kikuchi bands lose their intensity after the sample is cooled as shown in Figure 3(b), indicating degradation in surface crystal quality. One possible explanation for this phenomenon is the over-oxidation of the film surface due to the presence of residual oxygen and water in the chamber, which cannot be pumped out quickly enough after the growth. Water is also a likely by product of the decomposition of the precursor. A second possible reason may be due to adsorbed nitrogen and carbon on the film surface as a by-product from metal organics as shown in XPS spectra for uncapped sample in Supplemental Information S3. On the other hand, SNO samples show consistently improved RHEED images after capping with SHO, which can be found in supporting materials in Figure S1. This indicates high crystallinity of the SHO capping layer and better preservation of SNO from over-oxidation during cooldown. It should be noted that the medium-capped SNO sample shows relatively weak RHEED pattern after SHO capping as found in Figure S1 because of inconsistent Hf flux that produced a Hf-rich capping layer.

AFM measurements were performed to verify the surface morphology of the SNO films. Figure 3(c) shows the AFM topography of uncapped SNO film grown on GSO substrate with surface roughness of ~ 1.1 nm indicating the film is highly smooth. Additionally, AFM topography of SHO shows the surface roughness of ~2 nm as shown in Figure S2 indicating smooth capping surface. Figure 3(d) shows the XRD data for uncapped and SHO capped SNO films grown on GSO substrates. An uncapped SNO sample exhibits no clear evidence of crystalline perovskite film while capped samples have extra peak at lower 2θ. The intensity of an extra peak increases with increasing thickness of a capping, reflecting improved stoichiometry and crystallinity as predicted by the RHEED.

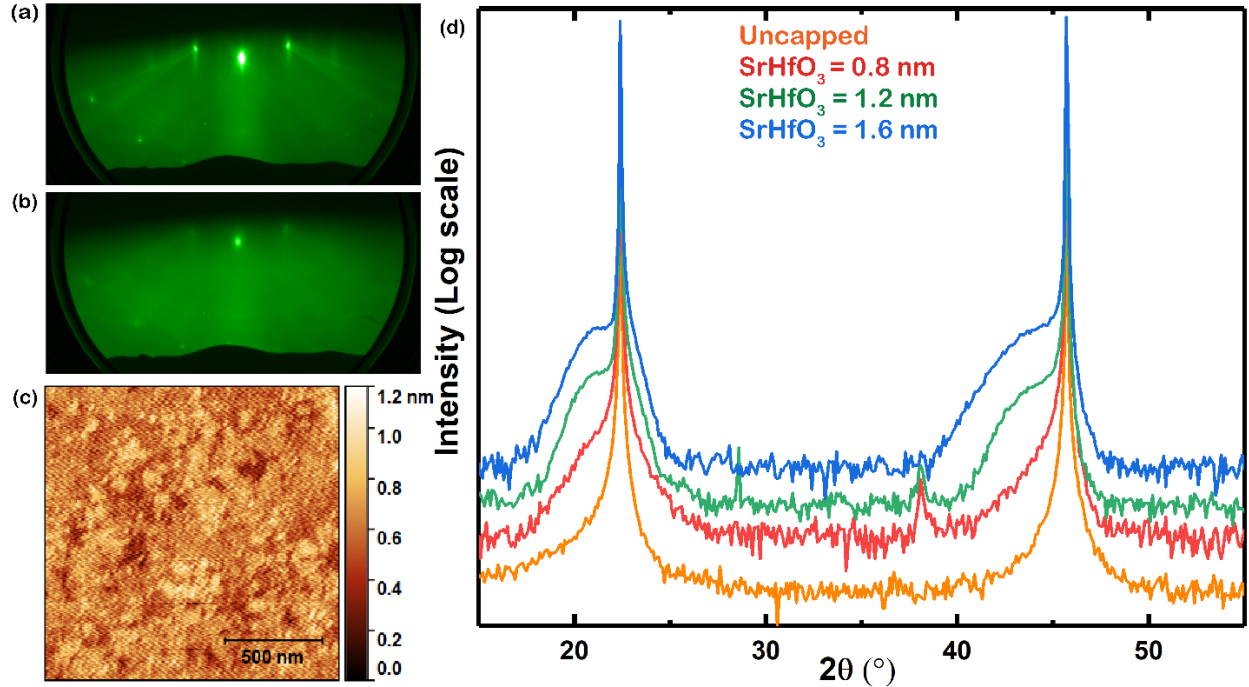

*Figure 3: RHEED image of SNO film (a) hot right after growth and (b) after cooling down on GSO along 110 azimuth, (c) AFM topography of uncapped SNO films, (d) XRD on uncapped and SHO capped samples*

To probe the role of capping on preservation of the $Nb^{4+}$ state, *in situ* XPS analysis of the Nb 3d core level on uncapped and SHO capped SNO films was performed. Figure 3(a-d) show the Nb 3d core level deconvolution for uncapped and SHO capped SNO films using constraints from table S1 as provided in supplemental information. As the Hf 4d core level significantly overlaps with the Nb 3d core level, the peaks must be carefully deconvoluted to determine the ratio of $Nb^{4+}$ to $Nb^{5+}$. A clear extra peak at lower binding energy in Nb 3d is significantly increased in capped samples compared to an uncapped sample that reflects the SHO capping is highly effective in preserving the $Nb^{4+}$ state. This peak has previously been observed after sputter cleaning of $NbO_2$ after atmospheric exposure [36] and *in situ* for high-quality $NbO_2$ films grown by MBE using an electron-beam evaporation source [16]. *In situ* measurements of uncapped $SrNbO_3$ films grown by PLD showed significantly smaller concentrations of $Nb^{4+}$ by comparison [14], indicating that the $SrHfO_3$ capping layer is important even for *in situ* studies of the material. Explicit quantification of the ratio of $Nb^{4+}$ to $Nb^{5+}$ is complicated by the satellite feature that is expected even for an ideal system with only $d^1$ states for all Nb cations [16]. However, it is clear that some $Nb^{5+}$ is still present even in the capped samples, leading to a chemical formula of $SrNbO_{3+\delta}$ for the material.

XPS valence band spectra were acquired simultaneously with the Nb 3d core level data to measure the density of states for the Nb 4d electronic states near the Fermi level. Figure 4(e) shows the valence band spectra for uncapped and SHO capped samples. The thickness and quality of the capping layer plays a clear role in the preservation of the single Nb 4d electron at the Fermi level, with the thickest capping layer providing the largest peak. The medium-capped sample shows relatively lower electron concentration compared to a thin capped sample as shown in Figure 4(e). That is likely related to the off-stoichiometry of the SHO capping layer governed by fluctuation in Hf flux during SHO deposition for that sample. This can be inferred from Figure 3(b) as we see a more intense Hf 4d core level compared to Nb 3d core level among all capped samples. This result can also be correlated with lowest quality of RHEED image for medium capped film as shown in middle panel of Figure S1. It is thus clear that the effectiveness of SHO as a capping layer for SNO is dependent on the crystalline quality of the SHO film.

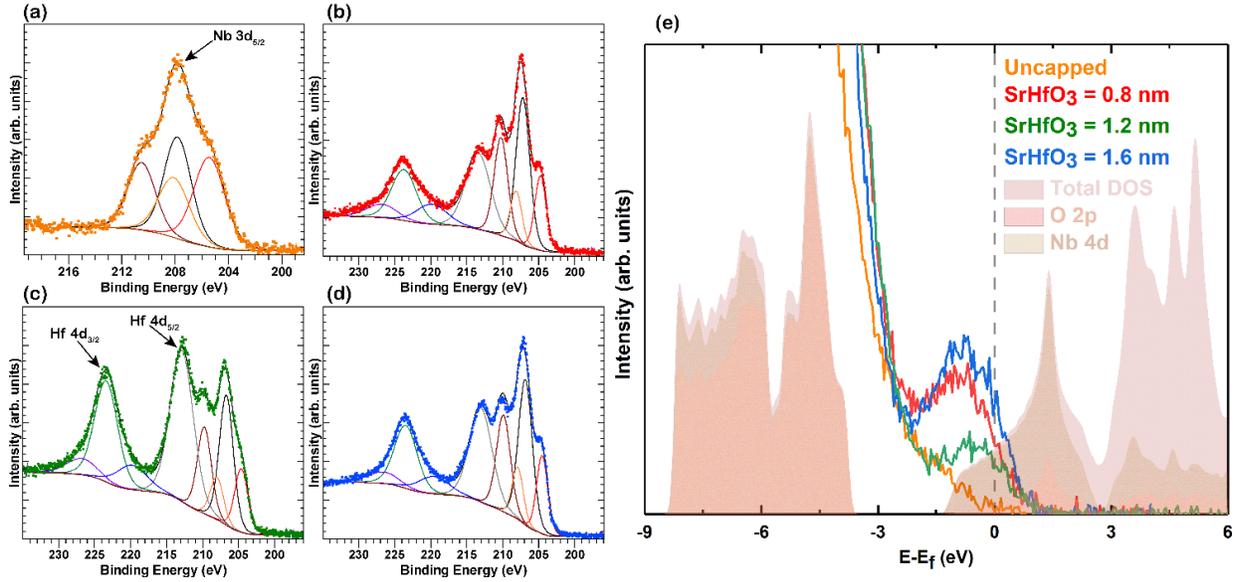

*Figure 4: Nb 3d core level deconvolution for SNO thin films capped with (a) uncapped, (b) thin (0.8 nm) capped, (c) medium (1.2 nm) capped, (d) thick (1.6 nm) capping of SHO. Fits to the data show spin-orbit split peaks of $Nb^{4+}$ (low binding energy) and $Nb^{5+}$ (high binding energy) features. (e) Valence band XPS data showing density of states near the Fermi level for all four samples. Partially transparent density of states from DFT model of $SrNbO_3$ is taken from Figure 2.*

XANES offers a more bulk-sensitive probe of the Nb valence after atmospheric exposure. Hence, to supplement the XPS analysis on oxidation state of Nb in the film, XANES analysis on Nb K-edge was also carried out. Figure 5 shows the XANES spectra on the Nb K-edge for SHO-capped samples with reference spectra for $Nb^{4+}$ ($NbO_2$) and $Nb^{5+}$ ($Nb_2O_5$) [37]. The Nb K-edge shows that the pre-edge features of the thinnest capped SNO sample mirrors $Nb_2O_5$ spectra, indicating more $Nb^{5+}$ state compared to other samples. The white line is also shifted to greater photon energy, consistent with a higher formal charge. The Nb K-edge energy in the inset of the figure shows that the thickest capped sample the spectrum white line is shifted to lower photon energy, indicating the greatest concentration of $Nb^{4+}$ state among all of the samples. This observation indicates that the capping has great role on preserving Nb oxidation state in SNO film after air exposure.

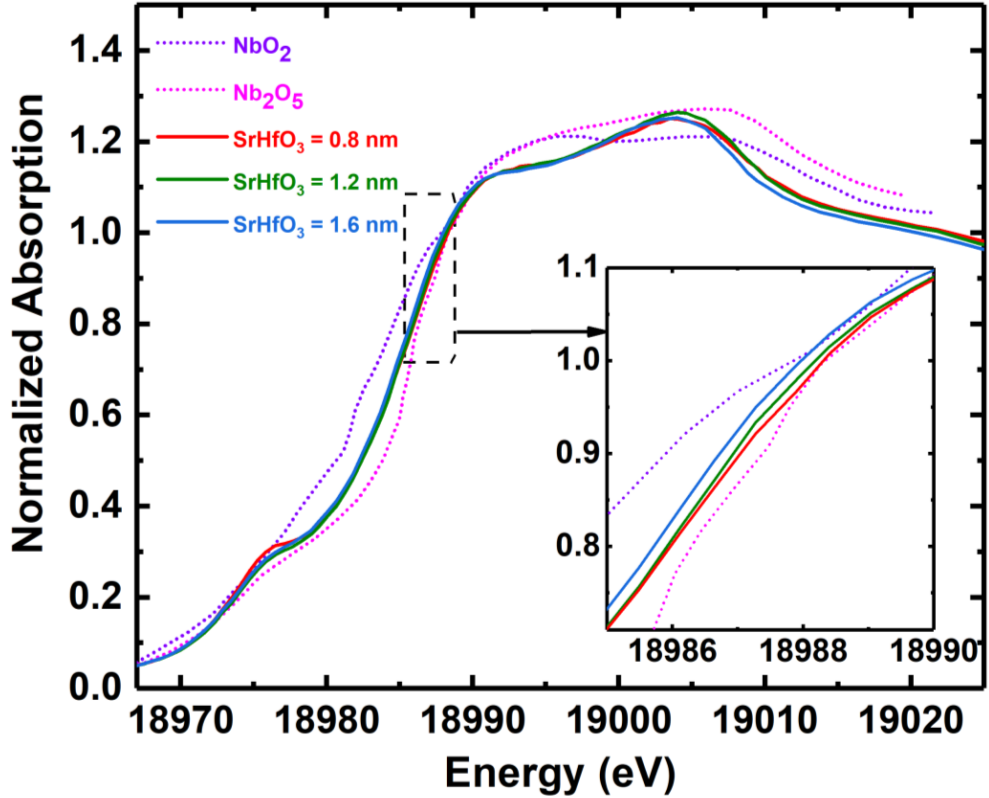

*Figure 5: Nb K edge XANES data for three SrHfO₃-capped samples. NbO$_2$ (Nb$^{4+}$) and Nb$_2$O$_5$ (Nb$^{5+}$) reference spectra were taken from Marini, et al. [37]. (Inset) High-resolution perspective of three capped samples showing enhanced Nb$^{4+}$ charge state in sample with 1.6 nm capping layer.*

**Discussion and Conclusion:**
Successful growth of metastable SNO by hMBE using the TDTBN precursor is performed for the first time. Good lattice match to the GSO substrate supported the growth of high quality epitaxial SNO films, which can be inferred from the high quality RHEED in Figure 2(a) and Figure S1. Degrading RHEED quality of an uncapped sample during cool down shows that residual oxygen in the chamber over-oxidizes the SNO film surface. Our observation shows that minimizing exposure to oxygen after film growth is critical to preserving a stable film surface, as shown in the RHEED images in the supplemental information and the valence band density of states in Figure 4. Over-oxidation of the SNO film surface can be reduced by depositing a dielectric caping layer such as SHO. Cooling down in vacuum for uncapped samples reduces the formation of competing phases such as amorphous Sr$_2$Nb$_2$O$_7$ [15]. However, surface-sensitive studies such as XPS and ARPES are still likely to detect the majority of the signal from the top few unit cells where Nb$^{5+}$ will be the dominant charge state. On the other hand, ex-situ AFM topography reveals that uncapped samples are highly smooth even after air exposure, suggesting that amorphization is fairly uniform across the sample. Thus, a smooth surface is not sufficient evidence to confirm ideal oxygen stoichiometry at the film surface. Our XRD results show that thin SNO films (less than 5 nm) had degraded crystal quality without a capping layer. Thus, we suggest that future studies on ultrathin SNO films should take care to decouple intrinsic phenomena related to the novel physics that can occur in the system from features that may occur due to variation in oxygen stoichiometry away from the ideal SrNbO$_3$ chemical formula.

In conclusion, our study of the synthesis of metastable SrNbO$_{3+\delta}$ thin films demonstrates the capabilities of the hMBE technique with a new TDTBN precursor. As predicted by our DFT calculations, the SNO films have valence band features at the Fermi level based on XPS analysis which is strongly dependent on the Nb charge state. The surface of the sample is preserved during cooldown by depositing a capping layer of SrHfO$_3$ immediately after SrNbO$_3$ growth, which increases the electronic density of states near the Fermi level. XRD, XAS and XPS used to characterize the SNO film and effectiveness of the SHO capping at various thicknesses indicate that both the quality and the thickness of the capping layer are important to preserving the surface. The ability to deposit SrNbO$_3$ films by hMBE opens the door for future exploration of interfacial phenomena in heterostructures, which has not been explored previously in films.

**Acknowledgements:**
ST and RBC gratefully acknowledge support from the Air Force Office of Scientific Research under award number FA9550-20-1-0034. This research used resources of the Advanced Photon Source, an Office of Science User Facility operated for the U.S. Department of Energy (DOE) Office of Science by Argonne National Laboratory, and was supported by the U.S. DOE under Contract No. DE-AC02-06CH11357.

Supplemental Information:

*Table S1: Position constraints for Nb3d core level deconvolution*

| Sample ID | Position Constraints (eV) | | | |
|---|---|---|---|---|
| | Nb 5+ Spin-orbit splitting (black/brown) | Nb 4+ Spin-orbit splitting (red/orange) | Hf 4+ (grey/green) | Hf satellite peaks relative to primary Hf$^{4+}$ (blue/violet) |
| Uncapped SrNbO$_3$ | 2.7 | 2.65 | - | - |
| Capped samples | 3.05 | 3.4 | Unconstrained | 7.0 |

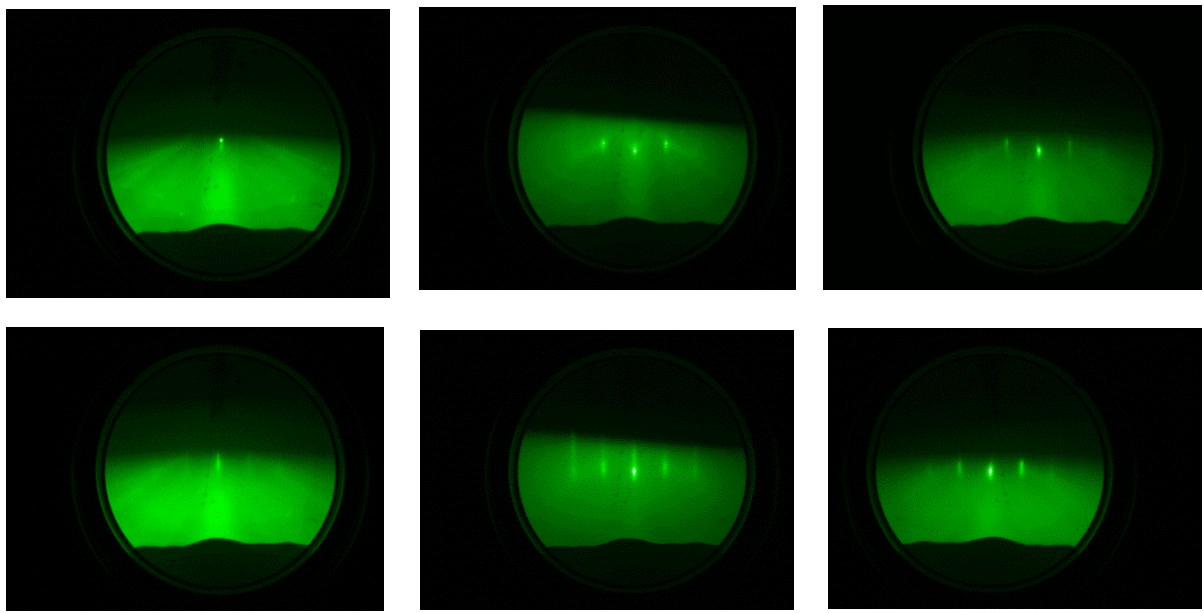

*Figure S1: RHEED images along 110 azimuths during hot for SrNbO$_3$ films before (top) and after (bottom) SrHfO$_3$ capping of (left) 2 unit cells, (middle) 3 unit cells, and (right) 4 unit cells.*

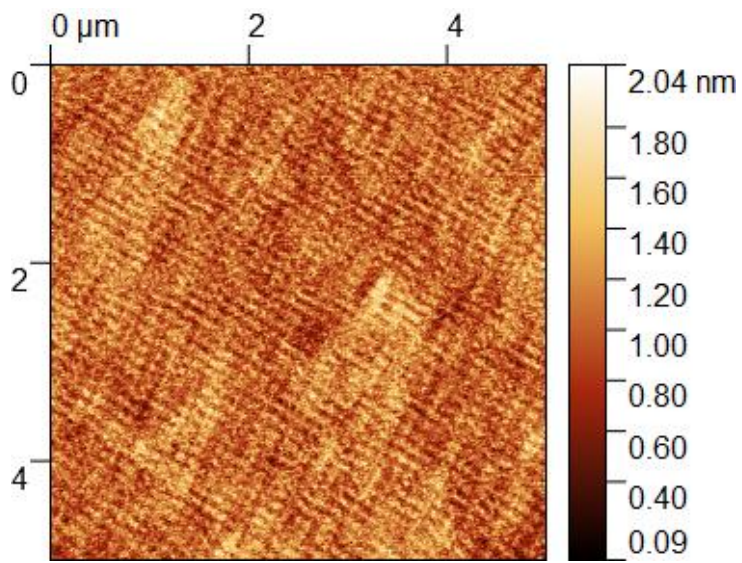

Figure S2: AFM topography of SrHfO$_3$ film

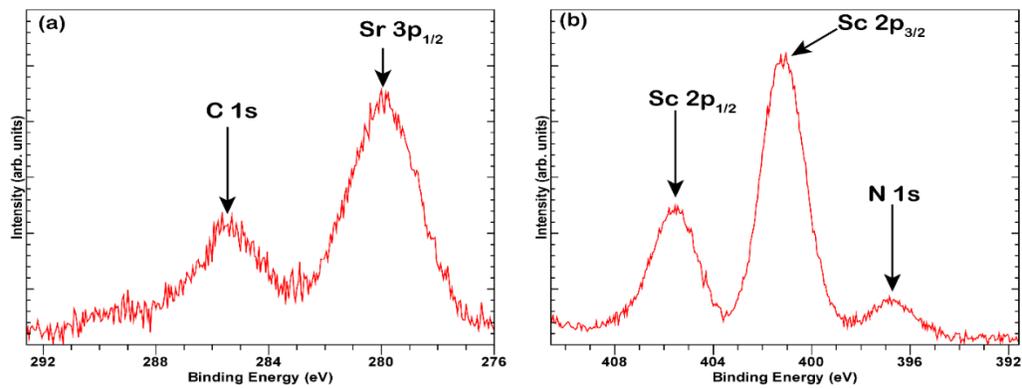

Figure S3: (a) C 1s and (b) N 1s on uncapped SNO sample

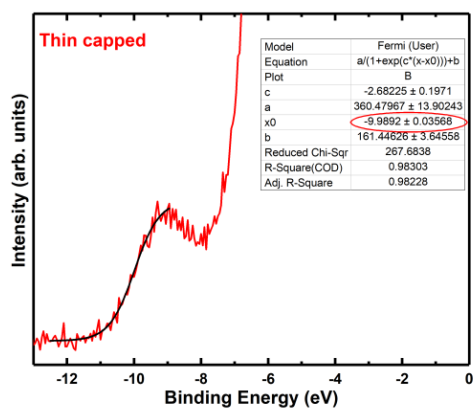

*Figure S4: XPS valence band spectra of a thin capped sample fitted with Fermi-Dirac function*

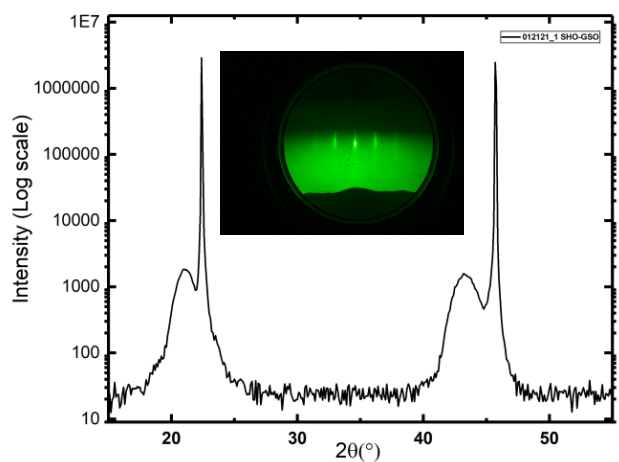

*Figure S5: XRD of SrHfO$_3$ on GdScO$_3$ and RHEED (inset)*